\documentclass[12pt]{iopart}

  \usepackage{latexsym,bm,amssymb,amsfonts}
  \usepackage{epsfig,graphics,graphicx}

  \newcommand{\beq}{\begin{eqnarray}}
  \newcommand{\eeq}{\end{eqnarray}}

\begin{document}

\title{Parton picture for the strongly coupled SYM plasma}

\author{Edmond Iancu}

\address{Institut de Physique Th\'eorique de Saclay,
 F-91191 Gif-sur-Yvette, France}
        \ead{edmond.iancu@cea.fr}

\begin{abstract} Deep inelastic scattering off the strongly coupled
${\mathcal N}=4$ supersymmetric Yang--Mills plasma at finite temperature
can be computed within the AdS/CFT correspondence, with results which are
suggestive of a parton picture for the plasma. Via successive branchings,
essentially all partons cascade down to very small values of the
longitudinal momentum fraction $x$ and to transverse momenta smaller than
the saturation momentum $Q_s\sim T/x$. This scale $Q_s$ controls the
plasma interactions with a hard probe, in particular, the jet energy loss
and its transverse momentum broadening.
\end{abstract}

\section{Introduction}

One of the most interesting suggestions emerging from the experimental
results at RHIC is that the deconfined, `quark--gluon', matter produced
in the early stages of an ultrarelativistic nucleus--nucleus collision
might be strongly interacting. This observation motivated a multitude of
applications of the AdS/CFT correspondence to problems involving a
strongly--coupled gauge plasma at finite temperature and/or finite quark
density. While early applications have focused on the long--range and
large--time properties of the plasma, so like hydrodynamics, more recent
studies have been also concerned with the response of the plasma to a
`hard probe' --- an energetic `quark' or `current' which probes the
structure of the plasma on space--time scales much shorter than the
characteristic thermal scale $1/T$ (with $T$ being the temperature).

From the experience with QCD one knows that the simplest hard probe is an
electromagnetic current. In deep inelastic scattering (DIS), the exchange
of a highly virtual space--like photon between a lepton and a hadron acts
as a probe of the hadron parton structure on the resolution scales set by
the process kinematics: if $Q^2$ is (minus) the photon virtuality and $s$
is the invariant photon--hadron energy squared, then the photon couples
to quark excitations having transverse momenta $k_\perp\lesssim Q$ and a
longitudinal momentum fraction $x\sim Q^2/s$. Also, the partonic
fluctuation of a space--like current can mimic a quark--antiquark
`meson', which is nearly on--shell in a frame in which the current has a
high energy. Furthermore, the decay of the time--like photon produced in
electron--positron annihilation is the simplest device to produce and
study hadronic jets in QCD. Thus, by studying the propagation of an
energetic current through the plasma one has access to quantities like
the plasma parton distributions, the meson screening length, or the
energy loss and the momentum broadening of a jet.

At strong coupling and large number of colors $N_c\gg 1$, the AdS/CFT
correspondence allows one to study the propagation of an Abelian
`${\mathcal R}$--current' through the finite--temperature plasma
described by the ${\mathcal N}=4$ supersymmetric Yang--Mills (SYM)
theory. (For recent reviews and more references see \cite{ADSREV}.) In
this context, DIS has been first addressed for the case of a dilaton
target, in Refs. \cite{Polchinski,HIM1}. These studies led to an
interesting picture for the partonic structure at strong coupling:
through successive branchings, all partons end up by `falling' below the
`saturation line', i.e., they occupy --- with occupation numbers of order
one --- the phase--space at transverse momenta below the saturation scale
$Q_s(x)$, which itself rises rapidly with $1/x$. Such a rapid increase,
which goes like $Q_s^2(x)\sim 1/x$ and hence is much faster than in
perturbative QCD, comes about because the high--energy scattering at
strong coupling is governed by a spin $j\simeq 2$ singularity
(corresponding to graviton exchange in the dual string theory), rather
than the usual $j\simeq 1$ singularity associated with the gluon exchange
at weak coupling.

In Refs. \cite{HIM2}, this partonic picture has been extended to a
finite--temperature SYM plasma  in the strong `t Hooft coupling limit
$\lambda\equiv g^2N_c\to \infty$ (meaning $N_c\to\infty$). The results of
these analyses will be briefly described in what follows.

\section{Deep inelastic scattering at strong coupling from AdS/CFT}

The strong coupling limit $\lambda\to \infty$ in the ${\mathcal N}=4$ SYM
gauge theory corresponds to the semiclassical, `supergravity',
approximation in the dual string theory, which lives in a
ten--dimensional curved space--time with metric $AdS_5\times S^5$. The
finite--temperature gauge plasma is `dual' to a black hole in $AdS_5$
which is homogeneous in the four Minkowski dimensions and whose AdS
radius $r_0$ is proportional to the temperature: $r_0=\pi R^2 T$, with
$R$ the curvature radius of $AdS_5$. The interaction between the
${\mathcal R}$--current $J_\mu$ and the plasma is then described as the
propagation of a massless vector field $A_\mu$ which obeys Maxwell
equations in the $AdS_5$ Schwarzschild geometry. The fundamental object
to be computed is the retarded current--current correlator,
   \beq
 \Pi_{\mu\nu}(q)\,\equiv\,i\int \rmd^4x\,\rme^{-iq\cdot x}\,\theta(x_0)\,
 \langle [J_\mu(x), J_\nu(0)]\rangle_T\,, \label{Rdef}  \eeq
whose imaginary part determines the cross--section for the current
interactions in the plasma, i.e., the plasma structure functions in the
{\em space--like} case $Q^2 \equiv -q^\mu q_\mu >0$ (`deep inelastic
scattering') and the rate for the current decay into `jets' in the {\em
time--like} case $Q^2<0$ (`$e^+e^-$ annihilation'). The imaginary part
arises in the supergravity calculation via the condition that the wave
$A_\mu$ has no reflected component returning from the horizon.
Physically, this means that the wave (current) can be absorbed by the
black hole (the plasma), but not also regenerated by the latter.

In what follows we shall focus on the space--like current, i.e., on the
problem of DIS off the plasma \cite{HIM2}. (The corresponding discussion
of a time--like current can be found in the second paper in Ref.
\cite{HIM2}; see also the related work in Ref. \cite{Karch}.) We choose
the current as a plane--wave propagating in the $z$ direction in the
plasma rest frame: $J_\mu(x)\,\propto \,\rme^{-i\omega t+iq z}$. Also, we
asume the high--energy and large--virtuality kinematics: $\omega\gg Q\gg
T$. The physical interpretation of the results can be facilitated by
choosing a different definition for the radial coordinate on $AdS_5$:
instead of $r$, it is preferable to work with the inverse coordinate
$\chi\equiv \pi R^2/r$, which via the UV/IR correspondence corresponds
(in the sense of being proportional) to the transverse size $L$ of the
partonic fluctuation of the current. Then, the $AdS_5$ boundary lies at
$\chi=0$ and the black--hole horizon at $\chi=1/T$.


The dynamics depends upon the competition between, on one hand, the
virtuality $Q^2$, which acts as a potential barrier preventing the
Maxwell wave $A_\mu$ to penetrate deeply inside $AdS_5$, and, on the
other hand, the product $\omega T^2$, which controls the strength of the
interactions between this wave and the black hole. (We recall that the
gravitational interactions are proportional to the energy density of the
two systems in interaction.) The relevant dimensionless parameter is
$Q^3/\omega T^2$, which can be also rewritten as $xQ/T$, where $x\equiv
Q^2/2\omega T$ (the Bjorken variable for DIS) has the physical meaning of
the longitudinal momentum fraction of the plasma `parton' struck by the
current.

Specifically, in the high--$Q^2$ regime at $Q^3/\omega T^2 \gg 1$, or
$x\gg T/Q$, the interaction with the plasma is relatively weak and the
dynamics is almost the same as in the vacuum: the wave penetrates in
$AdS_5$ up to a maximal distance $\chi_0\sim 1/Q$ where it gets stuck
against the potential barrier. Physically, this means that the current
fluctuates into a pair of partons (say, a quark--antiquark `meson') with
transverse size $L\sim 1/Q$. At finite temperature, however, the
potential barrier has only a finite width --- it extends up to a finite
distance $\chi_1\sim (1/T)\sqrt{Q/\omega}$ ---, so there is a small, but
non--zero, probability for the wave to cross the barrier via tunnel
effect. Physically, this means that the plasma structure function at
large $x$ is non--vanishing, but extremely small (exponentially
suppressed) : $F_{2}(x,Q^2)\,\propto\,{xN^2_cQ^2}\,
\exp\{-(x/x_s)^{1/2}\}$ for $x \gg x_s\equiv T/Q$. In other terms, when
probing the plasma on a transverse resolution scale $Q^2$, one finds that
there are essentially no partons with momentum fraction $x$ larger than
$T/Q\ll 1$.

Where are the partons then ? To answer this question, let us explore
smaller values of Bjorken's $x$, by increasing the energy $\omega$ at
fixed $Q^2$ and $T$. Then the barrier shrinks and eventually disappears;
this happens when $\omega$ is large enough for $\chi_1\sim\chi_0$, a
condition which can be solved either for $x$ (thus yielding $x\sim
x_s=T/Q$), or for $Q$, in which case it yields the {\em plasma saturation
momentum} : $Q_s^2(x,T)\sim T^2/x^2$. For higher energies, meaning $x <
x_s$, the barrier has disappeared and the Maxwell wave can propagate all
the way down to the black hole, into which it eventually falls, along a
trajectory which coincides with the `trailing string' of a heavy quark
\cite{Drag}. Physically, this means that the current has completely
dissipated into the plasma. We interpret this dissipation as {\it
medium--induced branching} : the current fragments into partons via
successive branchings, with a splitting rate proportional to a power of
the temperature. This branching continues until the energy and the
virtuality of the partons degrade down to values of order $T$. The
lifetime of the current (estimated as the duration of the fall of the
Maxwell wave into the black hole) is found as $\Delta t\sim \omega/Q_s^2
\propto \omega^{1/3}$ --- a result which agrees with a recent estimate of
the `gluon' lifetime in Ref. \cite{GubserGluon}. Since the current is
tantamount to a `meson' with size $1/Q$ and rapidity $\gamma=\omega/Q$,
our analysis also implies an upper limit on the transverse size of this
`meson' before it melts in the plasma: $L_{\rm max}\sim 1/Q_s \sim
1/\sqrt{\gamma}\,T$. This limit is consistent with the meson screening
length computed in Refs. \cite{Meson}. The saturation momentum $Q_s$
turns out to also be the scale which controls the energy loss
\cite{Drag,HIM2} and the transverse momentum broadening \cite{Broad,QSAT}
of a parton moving into the plasma. For instance, the rate for the energy
loss of a heavy quark reads (in the ultrarelativistic limit $\gamma\gg
1$) \cite{HIM2,QSAT}
 \beq\label{dragf}
 -\frac{\rmd \omega}{\rmd t}\,\sim\,\sqrt{\lambda}\,Q_s^2\,,\eeq
where one should keep in mind that the saturation scale in the r.h.s. is
itself a function of $\omega$, and hence of time: $Q_s^2 \sim (\omega
T^2)^{1/3}$. Eq.~(\ref{dragf}) may be viewed as the time--dependent
generalization of the `drag force' first computed in Refs. \cite{Drag}.

The complete absorbtion of the current by the plasma is tantamount to the
`black disk' limit for DIS: in this high--energy, or small--$x$, regime
the structure function is not only non--zero, but in fact it reaches its
maximal possible value allowed by unitarity. This value is found as
$F_{2}(x,Q^2)\,\sim\,{xN^2_cQ^2}$ for $x\sim x_s$, a result with a
natural physical interpretation: for a given resolution $Q^2$,
essentially all partons have momentum fractions $x\lesssim T/Q\ll 1$ and
occupation numbers $n\sim\mathcal{O}(1)$. This is similar to parton
saturation in pQCD, except that, now, the occupation numbers at
saturation are of order one, rather than being large ($n\sim 1/g^2 N_c$),
as it was the case at weak coupling.

This result has interesting consequences for a (hypothetic) high--energy
hadron--hadron collision, in which these partons would be liberated:
Since there are no partons carrying large longitudinal momenta, there
will be no `forward/backward jets' in the wake of the collision, that is,
no hadronic jets following the same directions of motion as the incoming
hadrons. Rather, all particles will be produced at central rapidities and
will be isotropically distributed in the transverse space. Similar
conclusions have been recently reached in Refs. \cite{Hofman}. This
picture looks quite different from that observed, say, in heavy ion
collisions at RHIC. Such a discrepancy suggests that much caution should
be taken when trying to extrapolate results from AdS/CFT to QCD.

\subsection*{Acknowledgments}
I would like to thank the Department of Energy's Institute for Nuclear
Theory at the University of Washington and the organizers of the INT
program ``From Strings to Things: String Theory Methods in QCD and Hadron
Physics'' for hospitality and partial support during the completion of
this work.

\end{document}